\documentclass[prd,showpacs,showkeys,nofootinbib,preprint]{revtex4-1}
\usepackage{amsmath,amsfonts,amssymb,color}
\usepackage{bm}
\usepackage{pgfplots}
\usepackage[colorlinks=true,urlcolor=blue,linkcolor=blue,citecolor=blue]{hyperref}

\begin{document} 

\title{General relativity as a special case of Poincar\'e gauge gravity}

\author{Yuri N. Obukhov}
\email{obukhov@ibrae.ac.ru}
\affiliation{Theoretical Physics Laboratory, Nuclear Safety Institute, 
Russian Academy of Sciences, B.Tulskaya 52, 115191 Moscow, Russia}

\author{Friedrich W. Hehl}
\email{hehl@thp.uni-koeln.de}
\affiliation{Institute for Theoretical Physics, University of Cologne, 50923 Cologne, Germany}

\begin{abstract}
We demonstrate that Einstein's general relativity theory arises as a special case in the framework of the Poincar\'e gauge theory of gravity under the assumption of a suitable nonminimal coupling of matter to the Riemann-Cartan geometry of spacetime. 
\end{abstract}

\pacs{04.20.Cv,04.50.Kd,04.50.-h}

\maketitle


\section{Introduction}\label{intro}

The Poincar\'e gauge [theory of] gravity (PG) arises as a natural extension of Einstein's general relativity theory (GR) by following gauge-theoretic principles, see \cite{Hehl:1976}-\cite{Feyn}. 

The Standard Model of fundamental particle physics is based on gauge theories for internal symmetries (described by the unitary groups $U(1), SU(2), SU(3)$). It clearly demonstrates that, apart from GR, the gauge idea underlies all physical theories of fundamental interactions. The geometrization of gravitational physics, by using the covariance and the equivalence principles, is similar to the geometrization of the three `physical interactions' (electromagnetic, weak and strong) by using the Yang-Mills type of approach. There is a difference, though, in that the Standard Model deals with fundamental symmetry groups acting in internal spaces, whereas gravity has to do with the symmetry of the external spacetime.

Fairly early there were attempts to understand gravity as a gauge theory. Utiyama \cite{Utiyama:1956} paved the way in this direction by using the Lorentz group $SO(1,3)$ as a gauge group for gravity. It turned out to be unsuccessful, though, since the current which couples to the Lorentz group is the angular momentum current. However, as we know from Newton's theory of gravity, it is the mass density or---according to special relativity---the energy-momentum current that gravity has as its source. The group of the local spacetime translations (related to diffeomorphisms) plays the central role in GR. This manifests itself in the well-known fact \cite{Feyn} that the gravitational field couples to the corresponding translational Noether current, namely the energy-momentum current (a.k.a.\ energy-momentum tensor). 

Accordingly, when constructing the gauge theory of gravity, it is necessary to investigate the conservation of the material energy-momentum current \cite{WeylBH} and the related invariance under {\it rigid} and, subsequently, under  {\it local} translations. The localization of the translational invariance then creates the gravitational field. As a result, already since the 1970s, a translational gauge theory (TG) was set up in the form of a teleparallelism theory \cite{Itin:2001,Hehl:2016glb,Itin:2017,Itin:2018}. The paper of Cho \cite{Cho:1975dh}, see also \cite{Nitsch:1979qn}, may be taken as a concise description of a translational gauge theory of gravity. Its structure is revisited from a modern geometrical point of view in the more recent papers \cite{Obukhov:2002tm,Pereira:2019}, see also \cite{Koivisto:2019,Delliou:2020}. For the technical details of the formalism of TG, one may refer to \cite{Aldrovandi:2013}.

As is well-known, fundamental particle physics is based on the Poincar\'e group, which is a semidirect product of the translation group with the Lorentz group. The fundamental particles are classified by mass and spin which arise in the representation theory of the Poincar\'e group. In accordance with the semidirect product structure of the Poincar\'e group, the Noether theorem gives rise to the two currents: the energy-momentum tensor (translational current) and the spin angular-momentum tensor (intrinsic rotational current); for a comprehensive review, see \cite{Hehl:1976,MAG,Blag:2002,Blag:2013,PBO:2017,Mielke:2017}. 

The resulting Poincar\'e gauge theory provides, in this gauge-theoretic framework, a natural extension of GR, with the energy-momentum and spin currents as the sources of the gravitational field \cite{Ivanenko:1983,Trautman:2006,yno:2006,yno:2018}. The spacetime is then characterized by a Riemann-Cartan geometry with non-vanishing {\it torsion} and non-vanishing {\it curvature}.

In this paper we demonstrate that GR can be consistently interpreted as a special case of Poincar\'e gauge gravity PG under the two crucial assumptions: (i) the PG Lagrangian has a certain {\it special} form, (ii) the matter couples {\it nonminimally} to the gravitational field of PG. This result is nontrivial for the following reason: The translational gauge theory of gravity TG, which is equivalent to GR, is, as such, applicable to spinless matter only. Here we clarify how to avoid this difficulty and we will include matter with spin angular momentum in a consistent way.

\section{Poincar\'e gauge gravity: formal structure}\label{PG}

Following the general Yang-Mills-Utiyama-Sciama-Kibble gauge-theoretic sche\-me, the 10-parameter Poincar\'e group $T_4\,\rtimes\,SO(1,3)$ gives rise to the 10-plet of the gauge potentials which are consistently identified with the components $e_i{}^\alpha$ of the orthonormal coframe $\vartheta^\alpha = e_i{}^\alpha dx^i$ (4 potentials corresponding to the translation subgroup $T_4$) and the components $\Gamma_{i}{}^{\alpha\beta} = -\,\Gamma_{i}{}^{\beta\alpha}$ of the Lorentz connection $\Gamma^{\alpha\beta} = \Gamma_{i}{}^{\alpha\beta} dx^i$ (6 potentials for the Lorentz subgroup $SO(1,3)$). The corresponding covariant curls, the field strengths of translations and Lorentz rotations, (\ref{Tor}) and (\ref{Cur}), are the two-forms of the {\it torsion} and the {\it curvature,} respectively. See Appendix \ref{appA} for the mathematical definitions.

Let us consider a generalization of the Einstein-Cartan model \cite{Trautman:2006} with a Lagrangian that contains all possible linear curvature invariants and all possible quadratic invariants of the torsion, as constructed from its irreducible parts (\ref{iT2})-(\ref{iT1}):
\begin{eqnarray}
V = \frac{1}{2\kappa c}\left\{\left(\eta_{\alpha\beta} + \overline{a}{}_0\vartheta_\alpha
\wedge\vartheta_\beta\right)\wedge R^{\alpha\beta} - 2\lambda_0\eta
- T^\alpha\wedge\sum_{I=1}^3\left[a_I\,{}^*({}^{(I)}T_\alpha)
+ \overline{a}_I\,{}^{(I)}T_\alpha\right]\right\}.\label{LQT}
\end{eqnarray}
For completeness, we included a term carrying the  cosmological constant $\lambda_0$. As compared to the Einstein-Cartan model, the new Lagrangian contains 6 additional (dimensionless) coupling constants: $\overline{a}_0;a_1, a_2, a_3$ and $\overline{a}_1, \overline{a}_2 = \overline{a}_3$. The two latter constants are equal because the two last terms in (\ref{LQT}) are the same,
\begin{equation}\label{T23}
T^\alpha\wedge{}^{(2)}T_\alpha = T^\alpha\wedge{}^{(3)}T_\alpha = {}^{(2)}T^\alpha\wedge{}^{(3)}T_\alpha,
\end{equation}
whereas $T^\alpha\wedge{}^{(1)}T_\alpha = {}^{(1)}T^\alpha\wedge{}^{(1)}T_\alpha$. One can prove these relations directly from the definitions (\ref{iT2})-(\ref{iT1}).

For the Lagrangian (\ref{LQT}) we find the variational derivatives
\begin{align}
{\mathcal E}_\alpha &:= {\frac{\delta V}{\delta\vartheta^{\alpha}}} = 
-\,DH_{\alpha} + E_{\alpha}, \label{dVt}\\ 
{\mathcal C}_{\alpha\beta} &:= {\frac{\delta V}{\delta\Gamma^{\alpha\beta}}} 
= -\,DH_{\alpha\beta} + E_{\alpha\beta}\,.\label{dVG}
\end{align}
Here we denoted as usual
\begin{eqnarray}
H_\alpha &=& -\,{\frac{\partial V}{\partial T^{\alpha}}} = {\frac 1{\kappa c}}\sum_{I=1}^3\left[a_I\,{}^*({}^{(I)}T_\alpha) 
+ \overline{a}_I\,{}^{(I)}T_\alpha\right], \label{HaQT}\\
H_{\alpha\beta} &=& -\,{\frac{\partial V}{\partial R^{\alpha\beta}}} = -\,{\frac {1}{2\kappa c}}
\left(\eta_{\alpha\beta} + \overline{a}{}_0\vartheta_\alpha\wedge\vartheta_\beta\right),\label{HabQT}\\
E_\alpha &=& {\frac{\partial V}{\partial\vartheta^{\alpha}}} = {\frac {1}{2\kappa c}}\left(\eta_{\alpha\beta\gamma}\wedge R^{\beta\gamma} + 
2\overline{a}{}_0R_{\alpha\beta}\wedge\vartheta^\beta - 2\lambda_0\eta_\alpha\right)\nonumber\\
&& + \,{\frac 12}\left[(e_\alpha\rfloor T^\beta)\wedge H_\beta - T^\beta\wedge 
e_\alpha\rfloor H_\beta\right],\label{EaQT}\\
E_{\alpha\beta} &=& {\frac{\partial V}{\partial\Gamma^{\alpha\beta}}} =
{\frac 12}\left(H_\alpha\wedge\vartheta_\beta - H_\beta\wedge\vartheta_\alpha\right).\label{EabQT}
\end{eqnarray}
The corresponding field equations of PG are derived from the variation of the total Lagrangian $V + L$ with respect to the Poincar\'e gauge potentials $\vartheta^\alpha$ and $\Gamma^{\alpha\beta}$:
\begin{eqnarray}
{\frac 12}\eta_{\alpha\beta\gamma}\wedge R^{\beta\gamma} + \overline{a}{}_0R_{\alpha\beta}
\wedge\vartheta^\beta - \lambda_0\eta_\alpha
- Dh_\alpha + q^{(T)}_\alpha &=& \kappa\,{\mathfrak T}_\alpha,\label{EQT1}\\
\eta_{\alpha\beta\gamma}\wedge T^{\gamma} + \overline{a}{}_0\left(T_\alpha\wedge\vartheta_\beta
- T_\beta\wedge\vartheta_\alpha \right) + h_\alpha\wedge\vartheta_\beta - h_\beta\wedge\vartheta_\alpha
&=& \kappa c\,{\mathfrak S}_{\alpha\beta}.\label{EQT2}
\end{eqnarray}
 Here we denoted the linear and the quadratic functions of the torsion as
\begin{eqnarray}
h_\alpha &:=& \kappa cH_\alpha = \sum_{I=1}^3\left[a_I\,{}^*({}^{(I)}T_\alpha) 
+ \overline{a}_I\,{}^{(I)}T_\alpha\right],\label{HlaQT}\\
q^{(T)}_\alpha &:=& {\frac 12}\left[(e_\alpha\rfloor T^\beta)\wedge h_\beta - T^\beta\wedge 
e_\alpha\rfloor h_\beta\right].\label{qa}
\end{eqnarray}
It is straightforward to prove the simple properties of these objects which follow directly from their definitions:
\begin{eqnarray}
\vartheta^\alpha\wedge q^{(T)}_\alpha &=& 0,\label{vtqT}\\
\vartheta^\alpha\wedge h_\alpha &=& -\,a_2{}^*T + \overline{a}_3{}^*\overline{T},\label{vth}\\
e^\alpha\rfloor h_\alpha &=& a_3 \overline{T} + \overline{a}_2 T.\label{eth}
\end{eqnarray}
An important technical remark is in order: the two-form (\ref{HlaQT}) and the three-form (\ref{qa}) satisfy the geometrical identity
\begin{equation}
h_\alpha\wedge T_\beta - h_\beta\wedge T_\alpha + q^{(T)}_\alpha\wedge\vartheta_\beta - 
q^{(T)}_\beta\wedge\vartheta_\alpha \equiv 0.\label{hTq}
\end{equation}
To verify this, we notice that $h_\alpha$ is a linear combination of the irreducible parts of the torsion and its dual, and use the identities (\ref{TQI})-(\ref{TQ2}). The relation (\ref{hTq}) is always valid irrespectively whether the field equations are fulfilled or not. 

The matter sources on the right-hand sides of the gravitational field equations (\ref{EQT1}) and (\ref{EQT2}) are the three-forms of the canonical energy-momentum current and the spin current of matter, respectively:
\begin{align}
{\mathfrak T}_\alpha &:= {\frac{\delta L}{\delta\vartheta^{\alpha}}}, \label{EM}\\ 
{\mathfrak S}_{\alpha\beta} &:= {\frac{\delta L}{\delta\Gamma^{\alpha\beta}}}.\label{S}
\end{align}

Up to this point, we have presented a general formalism and now we will specify the structure of the PG field Lagrangian (\ref{LQT}).

\section{Model Lagrangian and field equations}\label{FE}

The geometric identities (\ref{ID1}) and (\ref{Keta}) between the contortion one-form $K^{\mu\nu}$ and the torsion two-form $T^\alpha$ underlie the subsequent discussion.

Let us consider the Poincar\'e gauge model belonging to the class (\ref{LQT}) and characterized by the following coupling constants:
\begin{equation}\label{V0}
\left.\begin{split}
a_1 = -\,1,\quad a_2 = 2,\quad a_3 = {\frac 12},\\
\overline{a}_1 = -\,\overline{a}_0,\quad \overline{a}_2 =
-\,\overline{a}_0,\quad \overline{a}_3 = -\,\overline{a}_0.
\end{split}\right\}
\end{equation}
Here we will show that the Poincar\'e gauge model (\ref{V0}) is actually Einstein's general relativity theory (GR), provided the matter Lagrangian $L = L(\psi^A, d\psi^A, \vartheta^\alpha, \Gamma^{\alpha\beta}, T^\alpha)$ is {\it non-minimally} coupled to the matter fields $\psi^A$ by means of the Poincar\'e gauge potentials $\vartheta^\alpha, \Gamma^{\alpha\beta}$ {\it and} the torsion $T^\alpha$.

Before we continue with our calculations, let us have a look at the explicit form of our Lagrangian. Substituting (\ref{V0}) into (\ref{LQT}), we find
\begin{eqnarray}\label{gravLagr}
V &= &\frac{1}{\kappa c}[\eta_{\alpha\beta}\wedge R^{\alpha\beta}-2\lambda_0\eta \nonumber \\ &&
-\,T^\alpha \wedge^*(-  ^{(1)}T_\alpha+2^{(2)}T_\alpha +\frac{1}{2}\,^{(3)}T_\alpha)\nonumber\\
&& +\overline{a}_0\underbrace{(\vartheta_{\alpha}\wedge\vartheta_\beta
\wedge R^{\alpha\beta}+T^\alpha \wedge T_\alpha) }_{=d(\vartheta_\alpha\wedge T^\alpha)}\,].
\end{eqnarray}
In the first line, we have the Einstein-Cartan Lagrangian including the cosmological term, in the second line we find the so-called viable set of torsion-square pieces of teleparallel gravity, and in the third line, which is parity odd, there features an exact form, that is, we have a boundary term. Here $\vartheta_\alpha\wedge T^\alpha$ is proportional to the translational Chern-Simons three-form of PG, see \cite{MAG}; its derivative yields the Nieh-Yan identity \cite{Nieh:1981,Nieh:2018}, see the underbraced expression in (\ref{gravLagr}).

Let us now return to (\ref{gravLagr}) and calculate the field equations explicitly. We begin by evaluating the torsion functions (\ref{HlaQT}) and (\ref{qa}). Specifically for the model (\ref{gravLagr}), we find:
\begin{eqnarray}
h_\alpha &=& h^{(0)}_\alpha - \overline{a}_0\,T_\alpha,\qquad q_\alpha = q^{(0)}_\alpha,\label{hq}\\
h^{(0)}_\alpha &=& -\,{\frac 12}K^{\mu\nu}\wedge\eta_{\alpha\mu\nu},\label{h0a}\\
q^{(0)}_\alpha &=& K_\alpha{}^\beta\wedge h^{(0)}_\beta
+ {\frac 12}K_\gamma{}^\mu\wedge K^{\nu\gamma}\wedge\eta_{\alpha\mu\nu}.\label{q0a}
\end{eqnarray}
With the superscript $^{(0)}$, we denote all objects which refer to the parity-even sector of the model (\ref{gravLagr})---the first two lines in (\ref{gravLagr}).

The proof of (\ref{h0a}) is straightforward: one should combine the definition (\ref{HlaQT}) with the identity (\ref{ID1}). To verify (\ref{q0a}), we start with the definition of  $q^{(0)}_\alpha$, see (\ref{HlaQT}),
\begin{equation}
  q^{(0)}_\alpha = {\frac 12}\left[(e_\alpha\rfloor T^\beta)\wedge h^{(0)}_\beta
    - T^\beta\wedge e_\alpha\rfloor h^{(0)}_\beta\right]\label{q00},
\end{equation}
and evaluate the two terms on the right-hand side. Using (\ref{h0a}), we have
\begin{equation}
T^\beta\wedge e_\alpha\rfloor h^{(0)}_\beta = -\,{\frac 12}T^\beta\wedge\left\{(e_\alpha\rfloor
K^{\mu\nu})\eta_{\beta\mu\nu} + K^{\mu\nu}\eta_{\alpha\beta\mu\nu}\right\}.\label{Th0}
\end{equation}
For the first term we use another identity (\ref{Keta}) and we find
\begin{eqnarray}
T^\beta\wedge\eta_{\beta\mu\nu}(e_\alpha\rfloor K^{\mu\nu}) = -\,\vartheta_\mu\wedge K^{\rho\sigma}
\wedge\eta_{\nu\rho\sigma}(e_\alpha\rfloor K^{\mu\nu}) = (e_\alpha\rfloor T^\nu - K_\alpha{}^\nu)
\wedge K^{\rho\sigma}\wedge\eta_{\nu\rho\sigma},\label{Th1}
\end{eqnarray}
since $(e_\alpha\rfloor K^{\mu\nu})\vartheta_\mu = -\,e_\alpha\rfloor T^\nu + K_\alpha{}^\nu$. Consequently,
\begin{equation}
-\,{\frac 12}T^\beta\wedge\eta_{\beta\mu\nu}(e_\alpha\rfloor K^{\mu\nu}) = (e_\alpha\rfloor T^\beta)\wedge h^{(0)}_\beta - K_\alpha{}^\beta\wedge h^{(0)}_\beta,\label{Th2}
\end{equation}
and substituting this into (\ref{Th0}) and comparing it with (\ref{q00}), we derive
\begin{eqnarray}
q^{(0)}_\alpha = {\frac 12}K_\alpha{}^\beta\wedge h^{(0)}_\beta
+ {\frac 14}K^{\mu\nu}\wedge\eta_{\alpha\beta\mu\nu}T^\beta.\label{Th3} 
\end{eqnarray}
We note that $\eta_{\alpha\beta\mu\nu}T^\beta = D\eta_{\alpha\mu\nu} = K_\alpha{}^\beta\wedge\eta_{\beta\mu\nu} + K_\mu{}^\beta\wedge\eta_{\alpha\beta\nu} + K_\nu{}^\beta\wedge\eta_{\alpha\mu\beta}$. Hence
\begin{equation} 
{\frac 14}K^{\mu\nu}\wedge\eta_{\alpha\beta\mu\nu}T^\beta = {\frac 12}K_\alpha{}^\beta\wedge h^{(0)}_\beta
+ {\frac 12}K_\gamma{}^\mu\wedge K^{\nu\gamma}\wedge\eta_{\alpha\mu\nu}.\label{Th4}
\end{equation}
After substituting this into (\ref{Th3}), the proof of (\ref{q0a}) is completed. Incidentally, $h_\alpha^{(0)}$ and $q_\alpha^{(0)}$ satisfy the identity
\begin{equation}
\vartheta_\alpha\wedge q_\beta^{(0)} - \vartheta_\beta\wedge q_\alpha^{(0)} \equiv
T_\alpha\wedge h_\beta^{(0)} - T_\beta\wedge h_\alpha^{(0)},\label{hTq0}
\end{equation}
which is the special case of the general identity (\ref{hTq}).

We are now in a position to analyse the left-hand sides of the field equations of PG. At first, we observe that
\begin{equation}\label{RDh}
\overline{a}{}_0R_{\alpha\beta}\wedge\vartheta^\beta - Dh_\alpha = -\,Dh^{(0)}_\alpha,
\end{equation}
making use of (\ref{hq}) and the Bianchi identity $R_{\alpha\beta}\wedge\vartheta^\beta + DT_\alpha = 0$. Next, we have $Dh^{(0)}_\alpha = \widetilde{D}h^{(0)}_\alpha + K_\alpha{}^\beta\wedge h^{(0)}_\beta$. Thus, with the help of (\ref{h0a}) and (\ref{q0a}), we obtain
\begin{equation}
-\,Dh^{(0)}_\alpha + q^{(0)}_\alpha = {\frac 12}(\widetilde{D}K^{\mu\nu} + 
K_\gamma{}^\mu\wedge K^{\nu\gamma})\wedge\eta_{\alpha\mu\nu}.\label{Dhq0}
\end{equation}
As a result, the two field equations (\ref{EQT1}) and (\ref{EQT2}) of PG are recast into
\begin{eqnarray}
{\frac 12}\eta_{\alpha\beta\gamma}\wedge\widetilde{R}{}^{\beta\gamma}
- \lambda_0\,\eta_\alpha &=& \kappa\,{\mathfrak T}_\alpha,\label{EQ01}\\
0 &=& \kappa c\,{\mathfrak S}_{\alpha\beta}.\label{EQ02}
\end{eqnarray}
The left-hand side of (\ref{EQ01}) reduces to the Riemannian Einstein two-form by combining the decomposition (\ref{RR}) with (\ref{Dhq0}). The left-hand side of (\ref{EQ02}) vanishes in view of
(\ref{h0a}) and the identity (\ref{Keta}).

After clarifying the left-hand sides of the PG field equations, in the next section we turn to the analyses of the right-hand sides.

\section{Coupling of gravity to matter}\label{matter}

To finalize the discussion of model (\ref{gravLagr}), we need to analyze the coupling of matter to gravity. At the first sight, the second field equation of PG (\ref{EQ02}) looks contradictory, because it apparently tells us that the spin current of matter is zero. However, this is only true if we assume that matter couples to gravity in accordance with the minimal coupling principle. In the latter case, the material Lagrangian is a function of matter fields $\psi^A$ and their covariant derivatives $D\psi^A$.

This apparent inconsistency can be avoided if we make the crucial assumption that the coupling of matter to gravity is {\it non-minimal} and the matter Lagrangian $L = L(\psi^A, D\psi^A, \vartheta^\alpha$, $T^\alpha)$ depends on the translational gauge field strength, the {\it torsion,} too. Moreover, such a non-minimal coupling is very special in the sense that the torsion enters the matter Lagrangian only in the combination
\begin{equation}\label{Dpsit}
\Phi^A := D\psi^A - {\frac 12}K^{\alpha\beta}\wedge(\rho_{\alpha\beta})^A{}_B\,\psi^B.
\end{equation}
Here $(\rho_{\alpha\beta})^A{}_B$ are the generators of the Lorentz algebra which determine the transformation of the matter field under the local Lorentz rotation of the coframe,
\begin{equation}\label{infP}
\delta\vartheta^\mu = \varepsilon(x)_\nu{}^\mu\vartheta^\nu,\qquad
\delta\psi^A = - \,{\frac 12}\varepsilon^{\alpha\beta}(\rho_{\alpha\beta})^A{}_B\psi^B,
\end{equation}
with the infinitesimal parameters $\varepsilon^{\alpha\beta} = -\, \varepsilon^{\beta\alpha}$. The Lagrange-Noether machinery for the nonminimal coupling case is well developed \cite{MAG,yno:2006,yno:2018}. It yields for the material sources of the Poincar\'e gauge field---the canonical energy-momentum and spin currents---a well known result:
\begin{eqnarray}
{\mathfrak T}_{\alpha} &=&  (e_\alpha\rfloor D\psi^A)\wedge
{\frac{\partial L}{\partial D\psi^A}} + (e_\alpha\rfloor\psi^A)\wedge
{\frac{\partial L}{\partial\psi^A}} - e_\alpha\rfloor L\nonumber\\
&& -\,D{\frac{\partial L}{\partial T^\alpha}} + (e_{\alpha}\rfloor T^\beta)\wedge
{\frac{\partial L}{\partial T^\beta}},\label{Tnon}\\
c{\mathfrak S}_{\alpha\beta} &=& (\rho_{\alpha\beta})^A{}_B\,\psi^B
\wedge{\frac {\partial L} {\partial (D\psi^A)}}\nonumber\\
&& -\,\vartheta_\alpha\wedge {\frac {\partial L}{\partial T^\beta}} +
\vartheta_\beta\wedge {\frac {\partial L}{\partial T_\alpha}}.\label{Snon}
\end{eqnarray}
The second lines in these two expressions account for the non-minimal coupling.

We identify the first line of (\ref{Snon}) with the canonical spin current three-form {\it defined under the assumption of the minimal coupling}
\begin{equation}
c\overset{\rm m}{\mathfrak S}{}_{\alpha\beta} := (\rho_{\alpha\beta})^A{}_B\,\psi^B\wedge{\frac {\partial L} {\partial D\psi^A}}=-c\overset{\rm m}{\mathfrak S}{}_{\beta\alpha}.\label{Smin}
\end{equation}
This three-form can be equivalently represented by the ``spin energy potential'' two-form  $\overset{\rm m}{\mu}_\alpha$ according to 
\begin{equation}
\overset{\rm m}{\mathfrak S}{}_{\alpha\beta} = \vartheta_\alpha\wedge
\overset{\rm m}{\mu}_\beta - \vartheta_\beta\wedge
\overset{\rm m}{\mu}_\alpha.\label{Spsi}
\end{equation}
Resolved with respect to $\overset{\rm m}{\mu}_\alpha$, we find 
\begin{equation}
\overset{\rm m}{\mu}_\alpha = -\,e^\beta\rfloor\overset{\rm m}{\mathfrak S}
{}_{\alpha\beta} + {\frac 14}\vartheta_\alpha\wedge e^\beta\rfloor e^\gamma\rfloor
\overset{\rm m}{\mathfrak S}{}_{\beta\gamma}.\label{psim}
\end{equation}

Now we insert (\ref{Snon}) into the second field equation (\ref{EQ02}) and resolve the latter to find
\begin{equation}\label{DLTS}
{\frac {\partial L}{\partial T_\alpha}} = c\overset{\rm m}{\mu}_\alpha.
\end{equation}

Eq.(\ref{Dpsit}) yields
\begin{equation} {\frac{\partial L}{\partial D\psi^A}} =
  {\frac{\partial L}{\partial \Phi^A}}, \qquad {\frac{\partial
      \Phi^A}{\partial \psi^B}} = -\,{\frac 12}K^{\alpha\beta}
  (\rho_{\alpha\beta})^A{}_B.\label{DLD}
\end{equation}
Making use of these relations together with (\ref{DLTS}), allows us to recast the energy-momentum current of matter (\ref{Tnon}) into
\begin{eqnarray} {\mathfrak T}_{\alpha} = \overset{\rm m}{\mathfrak T}{}_{\alpha}
- c\widetilde{D}\,\overset{\rm m}{\mu}_\alpha,\label{TT}
\end{eqnarray}
where 
\begin{eqnarray}
\overset{\rm m}{\mathfrak T}{}_{\alpha} = (e_\alpha\rfloor \Phi^A)\wedge
{\frac{\partial L}{\partial \Phi^A}}- e_\alpha\rfloor L 
+ (e_\alpha\rfloor\psi^A)\wedge\Bigl({\frac{\partial L}{\partial\psi^A}}
+ {\frac{\partial \Phi^B}{\partial \psi^A}}{\frac{\partial L}{\partial \Phi^B}}\Bigr).
\label{Tmin}
\end{eqnarray}
The final piece, which completes the puzzle, comes up when one recognizes, with the help of (\ref{GG}), that $\Phi^A = \widetilde{D}\psi^A$ is, in fact, the Riemannian covariant derivative. Then we identify (\ref{Tmin}) with the usual canonical energy-momentum current \cite{Hehl:RMP} computed under the assumption of minimal coupling. In components, ${\mathfrak T}_\alpha = {\mathfrak T}_\alpha{}^\mu\eta_\mu$ and ${\mathfrak S}_{\alpha\beta} = {\mathfrak S}_{\alpha\beta}{}^\mu\eta_\mu$. Thus, we have
\begin{equation} {\mathfrak T}_{\alpha}{}^\mu = \overset{\rm m}{\mathfrak T}{}_{\alpha}{}^\mu
+ {\frac c2}\,\widetilde{D}_\nu\Bigl(\overset{\rm m}{\mathfrak S}{}^{\mu\nu}{}_\alpha
+ \overset{\rm m}{\mathfrak S}{}^\mu{}_\alpha{}^\nu
+ \overset{\rm m}{\mathfrak S}{}_\alpha{}^{\nu\mu}\Bigr).\label{TTc}
\end{equation}
We immediately recognize in this expression the so-called metric energy-momentum current symmetrized by means of the Belinfante-Rosenfeld procedure.

Thus, we have verified, indeed, that the Poincar\'e gauge field equations (\ref{EQ01})-(\ref{EQ02}) reproduce Einstein's GR for the Lagrangian (\ref{gravLagr}).

\section{Our model's particle content}\label{spectrum}

The conclusions above can be strengthened by the study of the dynamical particle content of the PG model (\ref{gravLagr}). As a background, we assume a {\it torsionless spacetime of constant curvature} $\lambda$, that is,
\begin{eqnarray}
\widehat{D}\widehat{\vartheta}{}^\alpha &=& d\widehat{\vartheta}{}^\alpha
+ \widehat{\Gamma}{}_\beta{}^\alpha\wedge\widehat{\vartheta}{}^\beta = 0,\label{noT}\\
\widehat{R}{}^{\alpha\beta} &=& d\widehat{\Gamma}{}^{\alpha\beta} + \widehat{\Gamma}{}_\gamma{}^\beta
\wedge\widehat{\Gamma}{}^{\alpha\gamma} = \lambda\,\widehat{\vartheta}{}^\alpha\wedge
\widehat{\vartheta}{}^\beta.\label{RB}  
\end{eqnarray}
Let us split the PG gauge potentials into background and perturbations:
\begin{eqnarray}
\vartheta^\alpha &=& \widehat{\vartheta}{}^\alpha + \chi^\alpha,\label{vp}\\
\Gamma^{\alpha\beta} &=& \widehat{\Gamma}{}^{\alpha\beta} + \gamma^{\alpha\beta}.\label{gp}  
\end{eqnarray}
The particle spectrum of a general quadratic PG model on the Minkowski background was considered in \cite{Blagojevic:2018}. Inserting (\ref{vp}) and (\ref{gp}) into the definitions of the torsion and the curvature, we find the expansions
\begin{eqnarray}
T^\alpha &=& \widehat{D}\chi^\alpha + \gamma_\beta{}^\alpha\wedge\widehat{\vartheta}{}^\beta
+ \gamma_\beta{}^\alpha\wedge\chi^\beta,\label{Texp}\\
R^{\alpha\beta} &=& \lambda\,\widehat{\vartheta}{}^\alpha\wedge\widehat{\vartheta}{}^\beta +
\widehat{D}\gamma^{\alpha\beta} + \gamma_\gamma{}^\beta\wedge\gamma^{\alpha\gamma}.\label{Rexp}  
\end{eqnarray}
The expansions of the $\eta$-basis can be straightforwardly obtained by making use of (\ref{vp}). Up to the second order in perturbations, we find:
\begin{eqnarray}
\eta &=& \widehat{\eta} + \chi^\alpha\wedge\widehat{\eta}{}_\alpha
+ {\frac 12}\chi^\alpha\wedge\chi^\beta\wedge\widehat{\eta}{}_{\alpha\beta},\label{ee1}\\
\eta_\alpha &=& \widehat{\eta}{}_\alpha + \chi^\beta\wedge\widehat{\eta}{}_{\alpha\beta}
+ {\frac 12}\chi^\beta\wedge\chi^\gamma\wedge\widehat{\eta}{}_{\alpha\beta\gamma},\label{ee2}\\
\eta_{\alpha\beta} &=& \widehat{\eta}{}_{\alpha\beta} + \chi^\gamma\wedge
\widehat{\eta}{}_{\alpha\beta\gamma} + {\frac 12}\chi^\mu\wedge\chi^\nu
\,\widehat{\eta}{}_{\alpha\beta\mu\nu},\label{ee3}\\
\eta_{\alpha\beta\gamma} &=& \widehat{\eta}{}_{\alpha\beta\gamma}
+ \chi^\delta\,\widehat{\eta}{}_{\alpha\beta\gamma\delta}.\label{ee4}
\end{eqnarray}
Substituting (\ref{Texp})-(\ref{ee4}) into (\ref{LQT}) and taking into account (\ref{gravLagr}), we obtain the quadratic Lagrangian which determines the dynamics of the gravitational perturbations
\begin{eqnarray}
V &=& {\frac {3\lambda}{\kappa c}}\,\widehat{\eta}
+ {\frac {1}{2\kappa c}}\Bigl\{d\Bigl[\left(\widehat{\eta}_{\alpha\beta}
- \widehat{\eta}_{\alpha\beta\gamma}\wedge\chi^\gamma\right)\wedge\gamma^{\alpha\beta} 
+ \overline{a}_0\left(\widehat{\vartheta}_\alpha\wedge\widehat{\vartheta}_\beta
- \widehat{\vartheta}_\alpha\wedge\chi_\beta\right)\wedge\gamma^{\alpha\beta}\Bigr]\nonumber\\
&& + \,{\frac 12}N^{\mu\nu}\wedge\widehat{\eta}_{\alpha\mu\nu}\wedge\widehat{D}\chi^\alpha 
- 2\lambda\chi^\alpha\wedge\chi^\beta\wedge\widehat{\eta}_{\alpha\beta}\Bigr\}.\label{V0exp}
\end{eqnarray}
The cosmological constant fixes the value of the constant curvature of the background:
\begin{equation}
\lambda = {\frac {\lambda_0}{3}}.\label{lala}
\end{equation}
The one-form $N_{\mu\nu} = -\,N_{\nu\mu}$ is constructed in terms of the covariant derivatives of the translational perturbations. Namely, by definition,
\begin{equation}
N^\alpha{}_\beta\wedge\widehat{\vartheta}{}^\beta = \widehat{D}\chi^\alpha,\label{NDc}
\end{equation}
so that explicitly
\begin{equation}\label{Nab}
N_{\alpha\beta} = {\frac 12}\left(\widehat{e}_\alpha\rfloor\widehat{D}\chi_\beta
- \widehat{e}{}_\beta\rfloor\widehat{D}\chi_\alpha - \widehat{\vartheta}{}^\gamma
\,\widehat{e}{}_\alpha\rfloor\widehat{e}{}_\beta\rfloor\widehat{D}\chi_\gamma\right).
\end{equation}

As we see, the rotational (Lorentz) perturbation $\gamma_{\mu\nu}$ is non-dynamical: it contributes only to the total derivative in (\ref{V0exp}), and hence the corresponding field equation is trivial. This is perfectly consistent with our previous analysis which demonstrated the vanishing of the left-hand side of the second field equation (\ref{EQ02}).

The last line of the linearized Lagrangian (\ref{V0exp}) determines the dynamics of the translational perturbation one-form $\chi^\alpha$. The latter has a nontrivial skew-symmetric part which is conveniently described in terms of the two-form
\begin{equation}
\overline{\chi} := {\frac 12}\chi^\alpha\wedge\widehat{\vartheta}{}_\alpha.\label{chiA}
\end{equation}
Indeed, decomposing $\chi^\alpha = \chi_\beta{}^\alpha\widehat{\vartheta}{}^\beta$, we find $\overline{\chi} = {\frac 12}\chi_{[\alpha\beta]}\widehat{\vartheta}{}^\alpha\wedge\widehat{\vartheta}{}^\beta$.

The symmetric part of the translational perturbation is then defined as
\begin{equation}
\varphi^\alpha := \chi^\alpha + \widehat{e}{}^\alpha\rfloor\overline{\chi},\label{chiS}
\end{equation}
so that $\varphi^\alpha\wedge\widehat{\vartheta}{}_\alpha = 0$, and in components $\varphi^\alpha = \chi^{(\alpha\beta)}\widehat{\vartheta}{}_\beta$.

As a result, the one-form (\ref{Nab}) is recast into
\begin{equation}\label{Nab1}
N_{\alpha\beta} = \widehat{e}_\alpha\rfloor\widehat{D}\varphi_\beta
- \widehat{e}{}_\beta\rfloor\widehat{D}\varphi_\alpha + \widehat{D}\left(
\widehat{e}_\alpha\rfloor\widehat{e}_\beta\rfloor\overline{\chi}\right),
\end{equation}
and we find
\begin{equation}
{\frac 12}N^{\mu\nu}\wedge\widehat{\eta}_{\alpha\mu\nu} = (\widehat{e}{}^\mu\rfloor
\widehat{D}\varphi^\nu)\wedge\widehat{\eta}_{\alpha\mu\nu} -
\widehat{D}\,{}^*\!(\widehat{\vartheta}{}_\alpha\wedge\overline{\chi}).\label{Neta2}
\end{equation}
This yields
\begin{eqnarray}
{\frac 12}N^{\mu\nu}\wedge\widehat{\eta}_{\alpha\mu\nu}\wedge\widehat{D}\chi^\alpha =
(\widehat{e}{}^\mu\rfloor\widehat{D}\varphi^\nu)\wedge\widehat{\eta}_{\alpha\mu\nu}
\wedge\widehat{D}\chi^\alpha - \widehat{D}\,{}^*\!(\widehat{\vartheta}{}_\alpha\wedge
\overline{\chi})\wedge\widehat{D}\chi^\alpha.\label{Neta3}
\end{eqnarray}
The last term can be transformed into a total derivative
\begin{eqnarray}
-\,\widehat{D}\,{}^*\!(\widehat{\vartheta}{}_\alpha\wedge\overline{\chi})
\wedge\widehat{D}\chi^\alpha\overline{\chi})\wedge\chi^\alpha 
= -\,d\left\{\chi^\alpha\wedge\widehat{D}\,{}^*\!(\widehat{\vartheta}{}_\alpha\wedge
\overline{\chi})\right\} + 4\lambda\,\overline{\chi}\wedge{}^*\overline{\chi},\label{TD1}
\end{eqnarray}
by noticing that $\widehat{D}\widehat{D}\,{}^*\!(\widehat{\vartheta}{}_\alpha\wedge\overline{\chi})\wedge\chi^\alpha = -\, \widehat{R}_\alpha{}^\beta\wedge{}^*\!(\widehat{\vartheta}{}_\beta\wedge\overline{\chi})\wedge\chi^\alpha = 2\lambda (\widehat{\vartheta}{}^\beta\wedge\overline{\chi})\wedge{}^*\!(\widehat{\vartheta}{}_\beta\wedge\overline{\chi}) = 4\lambda\,\overline{\chi}\wedge{}^*\overline{\chi}$. Here we used (\ref{RB}) and the definition (\ref{chiA}).

With the help of (\ref{chiA}), we recast the last term in the Lagrangian (\ref{V0exp}) into
\begin{equation}
- \,2\lambda\chi^\alpha\wedge\chi^\beta\wedge\widehat{\eta}_{\alpha\beta} =
- \,2\lambda\varphi^\alpha\wedge\varphi^\beta\wedge\widehat{\eta}_{\alpha\beta}
- 4\lambda\,\overline{\chi}\wedge{}^*\overline{\chi},\label{chichi}
\end{equation}
and observe that the last terms in (\ref{TD1}) and (\ref{chichi}) cancel each other. 

Next, we analyse the first term on the right-hand side of (\ref{Neta3}). Substituting the decomposition of the translational perturbation $\chi^\alpha = \varphi^\alpha - e^\alpha\rfloor\overline{\chi}$ into the latter, we find
\begin{equation}
-\,(\widehat{e}{}^\mu\rfloor\widehat{D}\varphi^\nu)\wedge\widehat{\eta}_{\alpha\mu\nu}
\wedge\widehat{D}(\widehat{e}^\alpha\rfloor\overline{\chi}) 
= -\,d\left\{\varphi^\alpha\wedge\widehat{D}\,{}^*\!(\widehat{\vartheta}{}_\alpha\wedge
\overline{\chi})\right\}.\label{TD2}
\end{equation}

Collecting all the intermediate derivations, we use (\ref{Neta3})-(\ref{TD2}) to bring the Lagrangian (\ref{V0exp}) into the final form
\begin{eqnarray}
V &=& V^{\rm non} + V^{\rm dyn},\label{VV}\\ 
V^{\rm dyn} &=& {\frac {1}{2\kappa c}}\left\{(\widehat{e}{}^\mu\rfloor\widehat{D}
\varphi^\nu)\wedge\widehat{\eta}_{\alpha\mu\nu}\wedge\widehat{D}\varphi^\alpha
- 2\,\lambda\varphi^\alpha\wedge\varphi^\beta\wedge\widehat{\eta}_{\alpha\beta}\right\}.\label{Vphi}
\end{eqnarray}
The first term on the right-hand side of (\ref{VV}) is non-dynamical one,
\begin{eqnarray}
V^{\rm non} &=& {\frac {1}{2\kappa c}}\left\{6\lambda\widehat{\eta} + d\,U^{\rm non}\right\},\label{Vnon}\\
U^{\rm non} &=& \left[\widehat{\eta}_{\alpha\beta} - \widehat{\eta}_{\alpha\beta\gamma}\wedge\chi^\gamma
+ \overline{a}_0\,(\widehat{\vartheta}_\alpha\wedge\widehat{\vartheta}_\beta - \widehat{\vartheta}_\alpha
\wedge\chi_\beta)\right]\wedge\gamma^{\alpha\beta} - \left(\chi^\alpha + \varphi^\alpha\right)\wedge
\widehat{D}\,{}^*\!(\widehat{\vartheta}{}_\alpha\wedge\overline{\chi}).\label{Unon}
\end{eqnarray}
Consequently, the rotational (Lorentz) perturbation $\gamma^{\alpha\beta}$ and the skew-sym\-metric part $\overline{\chi}$ of the translational perturbation both contribute merely to the total divergence term (\ref{Vnon}) in the Lagrangian, and hence they are both non-dynamical. The symmetric translational perturbation $\varphi^\alpha$ represents the only dynamical degree of freedom. According to (\ref{Vphi}), it satisfies the linearized version of Einstein's field equation:
\begin{equation}
\widehat{D}(\widehat{e}{}^\mu\rfloor\widehat{D}\varphi^\nu)\wedge\widehat{\eta}_{\alpha\mu\nu}
- 2\lambda\,\widehat{\eta}_{\alpha\beta}\wedge\varphi^\beta = 0.\label{Ephi}
\end{equation}
It is convenient to introduce a two-form
\begin{equation}
{\mathcal F}^\alpha := \widehat{D}\varphi^\alpha + (\widehat{e}_\beta\rfloor\widehat{D}\varphi^\beta)
\wedge\widehat{\vartheta}{}^\alpha.\label{Fierz}
\end{equation}
This object can be called a {\it Fierz field}, see \cite{Novello:2002} and \cite{yno:2003}. One can straightforwardly verify that
\begin{equation}
(\widehat{e}{}^\mu\rfloor\widehat{D}\varphi^\nu)\wedge\widehat{\eta}_{\alpha\mu\nu}
= {}^*\!{\mathcal F}_\alpha,\label{DpF}
\end{equation}
so that the field equation (\ref{Ephi}) is recast into
\begin{equation}
\widehat{D}\,{}^*\!{\mathcal F}_\alpha 
- 2\lambda\,\widehat{\eta}_{\alpha\beta}\wedge\varphi^\beta = 0,\label{Ephi1}
\end{equation}
whereas the linearized Lagrangian (\ref{Vphi}) can be compactly rewritten as
\begin{equation}
V^{\rm dyn} = {\frac {1}{2\kappa c}}\left\{{}^*\!{\mathcal F}_\alpha\wedge\widehat{D}
\varphi^\alpha - 2\lambda\varphi^\alpha\wedge\varphi^\beta\wedge
\widehat{\eta}_{\alpha\beta}\right\}.\label{Vphi1}
\end{equation}
Note that the covariant derivatives of (\ref{Ephi}) and (\ref{Ephi1}) vanish identically. Indeed, from (\ref{Ephi1}) we have 
\begin{equation}
\widehat{\vartheta}{}^\alpha\wedge{}^*\!{\mathcal F}_\alpha = -\,2
(\widehat{e}{}^\mu\rfloor\widehat{D}\varphi^\nu)\wedge\widehat{\eta}_{\mu\nu},\label{tF}
\end{equation}
and hence 
\begin{eqnarray}
\widehat{D}\widehat{D}\,{}^*\!{\mathcal F}_\alpha &=& -\,\widehat{R}_\alpha{}^\beta\wedge
\,{}^*\!{\mathcal F}_\beta = -\,\lambda\widehat{\vartheta}_\alpha\wedge\widehat{\vartheta}^\beta
\wedge\,{}^*\!{\mathcal F}_\beta\nonumber\\
&=& 2\lambda\widehat{\vartheta}_\alpha\wedge(\widehat{e}{}^\mu\rfloor\widehat{D}
\varphi^\nu)\wedge\widehat{\eta}_{\mu\nu}\nonumber\\
&=& 2\lambda\widehat{D}\varphi^\nu\wedge\widehat{\eta}_{\alpha\nu},\label{DDF}
\end{eqnarray}
which cancels exactly the derivative of the second term in (\ref{Ephi}) and (\ref{Ephi1}), namely $\widehat{D}(-\,2\lambda\,\widehat{\eta}_{\alpha\beta}\wedge\varphi^\beta)$. 

On the other hand, by multiplying (\ref{Ephi}) and (\ref{Ephi1}) with the coframe $\widehat{\vartheta}^\alpha\wedge$, one obtains a nontrivial equation for the trace $\varphi = \widehat{e}_\alpha\rfloor\varphi^\alpha$ of the translational perturbation:
\begin{equation}
\widehat{D}(\widehat{e}{}^\mu\rfloor\widehat{D}\varphi^\nu)\wedge\widehat{\eta}_{\mu\nu}
+ 3\lambda\widehat{\eta}_\mu\wedge\varphi^\mu = 0.\label{EphiT}
\end{equation}
Here we used (\ref{tF}) and $\widehat{\vartheta}^\alpha\wedge\widehat{\eta}_{\alpha\beta} = 3\widehat{\eta}_\beta$. 

It is instructive to rewrite everything in components. Starting with $\varphi^\alpha = \varphi_\beta{}^\alpha\,\widehat{\vartheta}{}^\beta$ (recall that $\varphi_{\alpha\beta} = \varphi_{\beta\alpha}$), we have $\widehat{D}\varphi^\alpha = {\frac 12}(\widehat{D}_\mu\varphi_\nu{}^\alpha - \widehat{D}_\nu\varphi_\mu{}^\alpha)\widehat{\vartheta}^\mu\wedge\widehat{\vartheta}^\nu$. Then we find for the Fierz field ${\mathcal F}^\alpha = {\frac 12}{\mathcal F}_{\mu\nu}{}^\alpha\,\widehat{\vartheta}^\mu\wedge\widehat{\vartheta}^\nu$ the components
\begin{eqnarray}\label{FFab}
{\mathcal F}_{\mu\nu}{}^\alpha = \widehat{D}_\mu\varphi_\nu{}^\alpha - \widehat{D}_\nu
\varphi_\mu{}^\alpha + \delta^\alpha_\nu(\widehat{D}^\lambda\varphi_{\mu\lambda} - \widehat{D}_\mu\varphi)
- \delta^\alpha_\mu(\widehat{D}^\lambda\varphi_{\nu\lambda} - \widehat{D}_\nu\varphi).
\end{eqnarray}
Here the trace scalar $\varphi = \widehat{e}_\alpha\rfloor\varphi^\alpha = \varphi_\alpha{}^\alpha$. 

Accordingly, the four-form Lagrangian (\ref{Vphi}), (\ref{Vphi1}) reads
\begin{eqnarray}\label{Vphi2}
V^{\rm dyn} &=& {\frac {\widehat{\eta}}{2\kappa c}}\left\{{\mathcal F}^{\mu\nu}{}_\alpha
\widehat{D}_\mu\varphi_\nu{}^\alpha + 2\lambda(\varphi_{\alpha\beta}\varphi^{\alpha\beta}
- \varphi^2)\right\}\\
&=& {\frac {\widehat{\eta}}{2\kappa c}}\Bigl\{\widehat{D}_\mu\varphi_{\nu\lambda}
\widehat{D}^\mu\varphi^{\nu\lambda} - \widehat{D}_\mu\varphi_{\nu\lambda}
\widehat{D}^\nu\varphi^{\mu\lambda} \nonumber\\
&& -\,(\widehat{D}^\rho\varphi_{\mu\rho} - \widehat{D}_\mu\varphi)(\widehat{D}_\sigma
\varphi^{\mu\sigma} - \widehat{D}^\mu\varphi) 
+ 2\lambda(\varphi_{\alpha\beta}\varphi^{\alpha\beta} - \varphi^2)\Bigr\},\label{Vphi3}
\end{eqnarray}
and the field equation (\ref{Ephi}) and (\ref{Ephi1}) is recast into
\begin{equation}
\widehat{D}{}^\nu{\mathcal F}_{\mu\nu}{}^\alpha + 2\lambda\,(\varphi_\mu{}^\alpha - 
\varphi\delta_\mu^\alpha) = 0.\label{Ephi2}
\end{equation}
Similarly, the trace equation (\ref{EphiT}) reads explicitly as
\begin{equation}
\widehat{D}{}_\mu(\widehat{D}{}^\mu\varphi - \widehat{D}{}_\nu\varphi^{\mu\nu})
- 3\lambda\varphi = 0.\label{EphiT1}
\end{equation}

Finally, a useful observation is in order. After introducing
\begin{equation}
u_{\mu\nu} := \varphi_{\mu\nu} - {\frac 12}g_{\mu\nu}\varphi,\label{uphi}
\end{equation}
we recast the Fierz tensor (\ref{FFab}) into
\begin{equation}\label{Fuab}
{\mathcal F}_{\mu\nu}{}^\alpha = \widehat{D}_\mu u_\nu{}^\alpha
- \widehat{D}_\nu u_\mu{}^\alpha + \delta^\alpha_\nu\widehat{D}^\lambda u_{\mu\lambda}
- \delta^\alpha_\mu\widehat{D}^\lambda u_{\nu\lambda}.
\end{equation}

Summarizing, we conclude that the particle spectrum of the model (\ref{gravLagr}) contains only the spin-2 graviton mode propagating on the de Sitter background.

\section{Discussion and conclusion}\label{conc}

In this paper we demonstrated that GR can be consistently interpreted as a specific model of Poincar\'e gauge gravity under two crucial assumptions: (i) The PG Lagrangian has a certain {\it special} form, namely that given in Eq. (\ref{gravLagr}). (ii) The matter couples {\it nonminimally} to the gravitational field of PG in accordance with the substitution specified in Eq. (\ref{Dpsit}).

Similar studies of relations between PG, teleparallel gravity and GR and the analysis of the relevant physical sources were earlier done in \cite{Nester:1977,So:2003}. It should be noted that one can formally recover GR by using the Lagrange multipliers method in PG \cite{MAG,Chen:2015}. However, this is achieved by extending the space of PG variables with auxiliary fields which are alien to the gauge-theoretic approach. The fundamental novelty of our result is the demonstration that GR arises as a special model in the framework of the genuine Poincar\'e gauge gravity theory where the only dynamical variables are the coframe $\vartheta^\alpha$ and the local Lorentz connection $\Gamma^{\alpha\beta}$ (i.e., the translational and rotational gauge potentials) with no extra degrees of freedom added.

\begin{acknowledgments}
For YNO this work was partially supported by the Russian Foundation for Basic Research (Grant No. 18-02-40056-mega).
\end{acknowledgments}
\bigskip

\appendix

\section{Riemann-Cartan geometry: torsion, curvature, and $\eta$-forms}\label{appA}

Our basic notation and conventions are as follows: Greek indices $\alpha, \beta, \dots = 0, \dots, 3$, denote the anholonomic components (for example, of a coframe $\vartheta^\alpha$), while the Latin indices $i,j,\dots =0,\dots, 3$, label the holonomic components ($dx^i$, e.g.). The Minkowski metric is $g_{\alpha\beta} = {\rm diag}(+1,-1,-1,-1)$.

The gravitational field is described by the coframe $\vartheta^\alpha = e_i{}^\alpha dx^a$ and the Lorentz connection $\Gamma^{\alpha\beta} = \Gamma_i{}^{\alpha\beta} dx^i$ one-forms. The translational and rotational field strengths read
\begin{eqnarray}
T^\alpha &=& D\vartheta^\alpha = d\vartheta^\alpha +\Gamma_\beta{}^\alpha\wedge
\vartheta^\beta,\label{Tor}\\ \label{Cur}
R^{\alpha\beta} &=& d\Gamma^{\alpha\beta} + \Gamma_\gamma{}^\beta\wedge\Gamma^{\alpha\gamma}.
\end{eqnarray}

The Riemannian connection one-form $\widetilde{\Gamma}_\beta{}^\alpha$ is uniquely defined by means of the vanishing torsion condition $d\vartheta^\alpha + \widetilde{\Gamma}_\beta{}^\alpha \wedge\vartheta^\beta = 0$. One can decompose the Lorentz (a.k.a.\ Riemann-Cartan) connection 
\begin{equation}
\Gamma^{\alpha\beta} = \widetilde{\Gamma}^{\alpha\beta} - K^{\alpha\beta}\label{GG}
\end{equation}
into the Riemannian and the post-Riemannian parts. The contortion one-form $K^{\alpha\beta} = - \,K^{\beta\alpha}$ is algebraically related to the torsion:
\begin{equation}
T^\alpha = K^\alpha{}_\beta\wedge\vartheta^\beta.\label{TK}
\end{equation}
Explicitly, we have for the contortion one-form:
\begin{equation}\label{Kab}
K_{\alpha\beta} = {\frac 12}\left(e_\alpha\rfloor T_\beta - e_\beta\rfloor T_\alpha
- \vartheta^\gamma\,e_\alpha\rfloor e_\beta\rfloor T_\gamma\right).
\end{equation}
By substituting (\ref{GG}) into (\ref{Cur}), we can decompose the curvature two-form into Riemannian and post-Riemannian parts:
\begin{equation}
R^{\alpha\beta} = \widetilde{R}^{\alpha\beta} - \widetilde{D}K^{\alpha\beta} + 
K_\gamma{}^\beta\wedge K^{\alpha\gamma}.\label{RR}
\end{equation}
Hereafter the Riemannian objects and operators (constructed with the help of the Riemannian connection) are denoted by the tilde. 

Denoting the volume 4-form by $\eta$, we construct the $\eta$-basis in the space of exterior forms the help of the interior products as $\eta_{\alpha_1\dots\alpha_p}:= e_{\alpha_p}\rfloor\dots e_{\alpha_1}\rfloor\eta$, $p=1,\dots,4$. They are related to the coframe $\theta$-basis via the Hodge dual operator $^\star$, for example, $\eta_{\alpha\beta} = {}^\star\left(\vartheta_\alpha\wedge\vartheta_\beta\right)$.

Useful relations for the products of the coframes:
\begin{eqnarray}
\vartheta^\mu\wedge\eta_\alpha &=& \delta^\mu_\alpha\,\eta,\label{ve1}\\
\vartheta^\mu\wedge\vartheta^\nu\wedge\eta_{\alpha\beta} &=& (\delta^\mu_\alpha\delta^\nu_\beta
- \delta^\nu_\alpha\delta^\mu_\beta)\,\eta,\label{ve2}\\
\vartheta^\beta\wedge\eta_{\alpha\mu\nu} &=& \delta^\beta_\alpha\,
\eta_{\mu\nu} + \delta^\beta_\mu\,\eta_{\nu\alpha} + \delta^\beta_\nu\,
\eta_{\alpha\mu},\label{idve}\\
\vartheta^\mu\wedge\vartheta^\nu\wedge\eta_{\alpha\beta\gamma} &=& 2(\delta^{[\mu}_\alpha\delta^{\nu]}_\beta
\eta_\gamma + \delta^{[\mu}_\beta\delta^{\nu]}_\gamma\eta_\alpha + \delta^{[\mu}_\gamma\delta^{\nu]}_\alpha
\eta_\beta).\label{ve3}
\end{eqnarray}

\section{Irreducible decomposition of the torsion}\label{appB}

The torsion two-form can be decomposed into the three irreducible pieces, $T^{\alpha}={}^{(1)}T^{\alpha} + {}^{(2)}T^{\alpha} + {}^{(3)}T^{\alpha}$, where
\begin{eqnarray}
{}^{(2)}T^{\alpha} &=& {\frac 13}\vartheta^{\alpha}\wedge T,\label{iT2}\\
{}^{(3)}T^{\alpha}&=& -\,{\frac 13}{}^*(\vartheta^{\alpha}\wedge\overline{T}),\label{iT3}\\
{}^{(1)}T^{\alpha}&=& T^{\alpha}-{}^{(2)}T^{\alpha} - {}^{(3)}T^{\alpha}.
\label{iT1}
\end{eqnarray}
Here the one-forms of the {\it trace} $T$ and the {\it axial trace} $\overline{T}$ of the torsion $T^{\alpha} = {\frac 12}T_{\rho\sigma}{}^{\alpha}\,\vartheta^\rho\wedge\vartheta^\sigma$ are defined in terms of the torsion components as follows:
\begin{eqnarray}
T &:=& e_\nu\rfloor T^\nu = T_{\mu\nu}{}^{\mu}\vartheta^\nu,\label{traceT}\\
\overline{T} &:=& {}^*(T^{\nu}\wedge\vartheta_{\nu}) = {\frac 12}T_{\rho\sigma\mu}\eta^{\rho\sigma\mu\nu}
\,\vartheta_\nu.\label{traceTa}
\end{eqnarray}

\section{Key identities}\label{appC}

There are several useful relations for the irreducible torsion parts. In particular consider $T^{\mu} = {\frac 12}T_{\rho\sigma}{}^{\mu}\,\vartheta^\rho\wedge\vartheta^\sigma$  and multiply it by $\eta_{\alpha\beta\mu}$. With the help of (\ref{ve3}) we find
\begin{equation}\label{Teta2}
T^\mu\wedge\eta_{\alpha\beta\mu} = {\frac 12}T_{\rho\sigma}{}^\mu\,\vartheta^\rho\wedge\vartheta^\sigma
\wedge\eta_{\alpha\beta\mu} = (T_{\alpha\beta}{}^\mu - 3\,{}^{(2)}T_{\alpha\beta}{}^\mu)\wedge\eta_\mu.
\end{equation}
On the other hand, for the dual ${}^*T^{\alpha} = {\frac 12}T_{\rho\sigma}{}^{\alpha}\,\eta^{\rho\sigma}$ we immediately verify
\begin{equation}\label{Teta3}
{}^*T_\alpha\wedge\vartheta_\beta - {}^*T_\beta\wedge\vartheta_\alpha = 
(T_{\alpha\beta}{}^\mu - 3\,{}^{(3)}T_{\alpha\beta}{}^\mu)\wedge\eta_\mu.
\end{equation}
Applying (\ref{Teta2}) and (\ref{Teta3}) to the irreducible torsion parts, we obtain the identities
\begin{eqnarray}\label{TTe1}
{}^{(1)}T^\mu\wedge\eta_{\alpha\beta\mu} &=& 2\,{}^*({}^{(1)}T_{[\alpha})\wedge\vartheta_{\beta]},\\
{}^{(2)}T^\mu\wedge\eta_{\alpha\beta\mu} &=& -\,4{}^*({}^{(2)}T_{[\alpha})
\wedge\vartheta_{\beta]},\label{TTe2}\\
{}^{(3)}T^\mu\wedge\eta_{\alpha\beta\mu} &=& -\,{}^*({}^{(3)}T_{[\alpha})
\wedge\vartheta_{\beta]}.\label{TTe3}
\end{eqnarray}

Another identity expresses the contortion in terms of the duals of the irreducible parts of the torsion:
\begin{eqnarray}
{\frac 12}\,K^{\mu\nu}\wedge\eta_{\alpha\mu\nu} \equiv {}^*\Big({}^{(1)}T_\alpha 
- 2{}^{(2)}T_\alpha - {\frac 1 2}{}^{(3)}T_\alpha\Big).\label{ID1}
\end{eqnarray}
To prove this, we substitute $K^{\mu\nu} = {\frac 12}(e^\mu\rfloor T^\nu - e^\nu\rfloor T^\mu - \vartheta^\lambda e^\mu\rfloor e^\nu\rfloor T_\lambda)$ into the left-hand side of (\ref{ID1}) and find:
\begin{equation}
K^{\mu\nu}\wedge\eta_{\alpha\mu\nu} = (e^\mu\rfloor T^\nu)\wedge\eta_{\alpha\mu\nu} - {\frac 12}
\,(e^\mu\rfloor e^\nu\rfloor T_\beta)\,\vartheta^\beta\wedge\eta_{\alpha\mu\nu}.\label{Ket1}
\end{equation}
In order to evaluate the first term, we start with
\begin{eqnarray}
(e^\mu\rfloor T^\nu)\wedge\eta_{\mu\nu} = e^\mu\rfloor (T^\nu\wedge\eta_{\mu\nu}) =
e^\mu\rfloor (\eta_\mu\wedge T) 
= -\,\eta_\mu\,e^\mu\rfloor T = -\,{}^*(\vartheta_\mu\,e^\mu\rfloor T) = -\,{}^*T,
\end{eqnarray}
where we used the identity $0\equiv e_\nu\rfloor (T^\nu\wedge\eta_\mu)=T\wedge\eta_\mu + T^\nu\wedge\eta_{\mu\nu}$. Applying the interior product $e_\alpha\rfloor$, we find
\begin{equation}
(e_\alpha\rfloor e^\mu\rfloor T^\nu)\,\eta_{\mu\nu} - (e^\mu\rfloor T^\nu)
\wedge\eta_{\alpha\mu\nu} = -\,e_\alpha\rfloor{}^*T.
\end{equation}
Thus the first term on the right hand side of (\ref{Ket1}) reads
\begin{eqnarray}
(e^\mu\rfloor T^\nu)\wedge\eta_{\alpha\mu\nu} &\equiv& e_\alpha\rfloor{}^*T 
+ (e_\alpha\rfloor e^\mu\rfloor T^\nu)\,\eta_{\mu\nu} \nonumber\\
&=& {}^*(T\wedge\vartheta_\alpha) + {}^*(\vartheta_\mu\wedge\vartheta_\nu\,
e_\alpha\rfloor e^\mu\rfloor T^\nu) \nonumber\\
&=& {}^*(-\,\vartheta_\alpha\wedge T + T_\alpha
- e_\alpha\rfloor (\vartheta^\nu\wedge T_\nu)).\label{Ket2}
\end{eqnarray}
The second term on the right hand side of (\ref{Ket1}) is easily computed with the help of (\ref{idve}): 
\begin{equation}
-\,{\frac 12}(e^\mu\rfloor e^\nu\rfloor T_\beta)\vartheta^\beta\wedge\eta_{\alpha\mu\nu}
\equiv {}^*\big(-\,{\frac 12}\vartheta_\mu\wedge\vartheta_\nu\,e^\mu\rfloor 
e^\nu\rfloor T_\alpha + \vartheta_{[\nu}\wedge\vartheta_{\alpha]}\,e^\nu\rfloor T\big)
= {}^*(T_\alpha - \vartheta_\alpha\wedge T).\label{Ket3}
\end{equation}
Collecting (\ref{Ket2}) and (\ref{Ket3}), we find:
\begin{equation}
K^{\mu\nu}\wedge\eta_{\alpha\mu\nu}\equiv {}^*\big(2T_\alpha - 2\vartheta_\alpha
\wedge T - e_\alpha\rfloor (\vartheta^\nu\wedge T_\nu)\big).
\end{equation}
Substituting the definitions (\ref{iT2})-(\ref{iT1}), one proves the identity (\ref{ID1}).

Taking the sum of (\ref{TTe1})-(\ref{TTe3}) and making use of (\ref{ID1}), we obtain another identity:
\begin{equation}\label{Keta}
T^\gamma\wedge\eta_{\alpha\beta\gamma} + \vartheta_{[\alpha}\wedge K^{\mu\nu}\wedge\eta_{\beta]\mu\nu}\equiv 0.
\end{equation}
The relations (\ref{TTe1})-(\ref{ID1}) and (\ref{Keta}) are {\it linear} in the torsion components.

In addition, there exist other identities which are {\it quadratic} in the torsion components. They read as follows: 
\begin{equation}
{}^*({}^{(I)}T_{[\alpha})\wedge T_{\beta]} + {}^*({}^{(I)}T_{\gamma})\wedge\vartheta_{[\alpha}\wedge
e_{\beta]}\rfloor T^\gamma = 0.\label{TQI}
\end{equation}
These identities hold for all irreducible parts, $I = 1, 2, 3$. Besides that, there are 
similar (sort of ``dual'') relations
\begin{eqnarray}
{}^{(1)}T_{[\alpha}\wedge T_{\beta]} + {}^{(1)}T_{\gamma}\wedge\vartheta_{[\alpha}\wedge
e_{\beta]}\rfloor T^\gamma &=& 0,\label{TQ1}\\
({}^{(2)}T_{[\alpha} + {}^{(3)}T_{[\alpha})\wedge T_{\beta]} + ({}^{(2)}T_{\gamma} + {}^{(3)}T_{\gamma})
\wedge\vartheta_{[\alpha}\wedge e_{\beta]}\rfloor T^\gamma &=& 0.\label{TQ2}
\end{eqnarray}
To prove the relations (\ref{TQI})-(\ref{TQ2}), one should directly use the definitions 
(\ref{iT2})-(\ref{iT1}).

\end{document}